\documentclass[12pt]{article}
\usepackage{amssymb,epsf,graphicx}
\usepackage{amsmath,amssymb,bm,epsfig,epsf,graphicx,color,float,url} 
\usepackage{caption}
\usepackage{subcaption}
\usepackage{colortbl}
\usepackage[all]{xy}

\makeatletter
	
	\@addtoreset{equation}{section}

	\@addtoreset{figure}{section}

	\@addtoreset{table}{section}
\makeatother

\input epsf.sty
\newcommand{\ee}{{\eps\eta}}
\def\bra#1{\langle #1 |}

\def\ket#1{|#1 \rangle}

\newcommand{\VEV}[1]{\left\langle #1\right\rangle}

\def\l{\left}
\def\r{\right}
\def\res#1{\oint\! \frac{d#1}{2\pi i} \,}

\let\eps = \varepsilon

\def \bea {\begin{eqnarray}}
\def \eea {\end{eqnarray}}
\def \bdm {\begin{displaymath}}
\def \edm {\end{displaymath}}

\def\hsp{\hspace{-8.5pt}}

\def\W6J#1#2#3#4#5#6{\mbox{\scriptsize${\protect\begin{bmatrix} #1,&\hsp #2,&\hsp #3\\ #4,&\hsp #5,&\hsp #6 \end{bmatrix}}$}}

\newcommand{\calN}{{\cal N}}

\newcommand{\calF}{{\cal F}}
\newcommand{\calE}{{\cal E}}

\newcommand{\calG}{{\cal G}}

\newcommand{\calA}{{\cal A}}

\newcommand{\calT}{{\cal T}}
\newcommand{\calH}{{\cal H}}

\newcommand{\calI}{{\cal I}}
\newcommand{\CR}[2]{\left[#1,#2\right]}
\newcommand{\ACR}[2]{\left\{#1,#2\right\}}
\newcommand{\nn}{\nonumber}
\newcommand{\p}{\partial}
\newcommand{\wt}[1]{\widetilde{#1}}

\begin{document}

\centerline{\large \bf Winding number  for arbitrary integer value }
\vspace*{1.0ex}
\centerline{\large \bf in Cubic String Field Theory}
\vspace*{8.0ex}

\centerline{\large \rm Toshiko Kojita}

\vspace*{4.0ex}
\begin{center}

{\it Core of STEM, Nara Women's Univ.,\\
Higashimachi, Kitauoya, Nara-City, Nara, Japan}
\vspace*{2.0ex}

{E-mail: { toshikojita@gmail.com}}

\end{center}

\vspace*{6.0ex}

\centerline{\bf Abstract}
\bigskip
We have focused on the topological structure of Cubic string field theory (CSFT). 
From the similarity of action between CSFT and Chern-Simons (CS) theory in three dimensions, 
we have investigated the quantity $\calN=\pi^2/3\int (UQU^{-1})^3$, 
which is expected to be the counterpart of winding number in CS theory. 
In our previous research, it was reported that $\calN$ can only take a limited number of integer values due to the inevitable anomalies
in Okawa type solution.
To overcome this unsatisfactory results,
we evaluate $\calN$ and EOM against a solution itself, $\calT$, for more general 
class of pure gauge form solution written in $K,B$ and $c$ in this paper.
Then we obtain general formula of $\calN$ and $\calT$.
From this result, we show that there is an infinite number of solutions
that $\calN$ takes any integer value while keeping $\calT=0$. 
We also show the gauge invariant observable of these solutions take appropriate values.
Furthermore, we evaluate the integral form of the BRST-exact quantity as surface integral.

 \vfill \eject

\baselineskip=16pt

\tableofcontents

\setcounter{footnote}{0}
\newpage
\section{Introduction}
\setcounter{equation}{0}
\label{sec:Introduction}

As already pointed out in Witten's original paper of cubic string field theory(CSFT)\cite{Witten:1985cc}, 
it is an important property for CSFT that the action is not invariant under the finite gauge transformation and receives the global anomaly $\calN$, which is defined by
\begin{align}
\calN&=\frac{\pi^2}{3}\int\l( UQU^{-1}\r)^3.
\label{eq:calN}
\end{align}
It is well known that the action of CSFT has the same algebraic structure 
as the integral of the Chern-Simons (CS) 3 form
\cite{Taylor:2003gn,Okawa:2012ica}.
The counterpart of $\calN$ in CS theory is a topological quantity counting the winding number 
from the manifold to the gauge group $N_{CS}=\int_M\text{tr}(gdg^{-1})^3/24\pi^2$.
The winding number is invariant under the infinitesimal gauge transformation, and is quantized to an integer values.

One of the features of $N_{CS}$ is that it can be written in the integral form of total derivative.
To give an simple example for $SU(2)$ gauge group, 
we employ hedgehog anzats for $g(x)$ on $M=S^3$, $g(x)=\exp(i\pi f(r)\hat{\boldsymbol{x}}\cdot \tau)$ with 
$r=|\boldsymbol{x}|$ and $\hat{\boldsymbol{x}}=\boldsymbol{x}/r$,
\begin{align}
N_{CS}&=\int d^3x \,\nabla\cdot \frac{2\pi f-\sin 2\pi f}{8\pi^2 r^2}\hat{\boldsymbol{x}}=f(\infty)-f(0).
\label{eq:hedgehog}
\end{align}
For the regularity of $g(x)$ at origin and infinity, $\sin \pi f(0)$ and $\sin \pi f(\infty)$ should be zero,
hence, $N_{CS}$ becomes integer.
Note that $N_{CS}$ seems to be naively zero since manifold is compact,
however, non-zero contributions comes from the origin and infinity where 
the Lie algebra valued function $\phi(x)$ in $g(x)=e^{i\phi(x)}$ is singular.
This property also holds regardless of the details of $g (x)$.
A notable point illustrated in this example is that the winding number arises from singularities of $g(x)$,
and the integral form of the total derivative is easy to see this property.

Now, from the similarity between both theory,
it would be a natural question whether $\calN$ has corresponding property to winding number. 
Essentially, CSFT does not have concepts of manifolds, gauge groups, surface integral, etc.,
that is, there are no obvious corresponding definitions of the topological quantities in CSFT.
On the other hand, research on whether $\calN$ has properties that can be considered equivalent in some sense
to winding number $N_{CS}$ would play an essential role to 
acquire the ultimate goal of understanding topological nature of CSFT.
Recently, it was reported that the discontinuities of vector profile
does not appear in the study of the constant magnetic filed configuration on a torus constructed by OSFT solution
\cite{Ishibashi:2016xak}. 
This result seems to contradict the viewpoint of low energy, 
this is because discontinuity is indispensable for defining the gauge field at every point on the torus.
The authors of the paper suggest the 
non-locality of the SFT gauge transformation in terms of the target space may obscure the concept of the 
coordinate patches.
And they also state we still do not know the definition or framework of stringy topological quantities 
that characterize such configuration.
Therefore, there already exist to need to provide a obvious definition of topological quantities in CSFT.

In previous paper\cite{Hata:2011ke}, 
we have proved $\calN$ can be written in terms of the integral form of the 
BRST-exact quantity, which corresponding to the integral of the total derivative in CS theory,  
regardless of the detail of the solution. 
And as a first step to answering our question, we examined whether $\calN$ takes an integer value for
a restricted universal class of pure gauge type solutions called Okawa type\cite{Okawa:2006vm}.
This solution is described by three type of string fields $K,B$ and $c$.
The results is that $\calN$ depends only on the singularity at $K=0$ and $K=\infty$ of the solution,
and $\calN$ cannot take an any integer value other than $0, \pm 1, \pm 2$ 
due to the inevitable anomaly\cite{Hata:2012cy}.
We also proposed a proper regularization for the singularity of the solutions.
This regularization makes the equation of the motion non-trivial,
then we evaluated the inner product of the equation of motion against the solution itself,
which is called EOM in the strong sense, additionally.
We demonstrated that EOM in the strong sense is not zero when an anomaly appears in $\calN$.
Therefore, unfortunately, it was concluded that $\calN$ takes on a limited number of integers.

Given the above fact, \cite{Hata:2019dwu,Hata:2019ybw} investigated $\calN$ and EOM in the strong sense 
in a wide class of classical solutions rather than Okawa type.
This classical solution consists of an infinite number of string fields written by $K,B,c$,
and satisfy the reality condition.
The author provided how to construct a classical solution 
such that $\calN$ is an integer while vanishing EOM in the strong sense. 
And the author actually constructed classical solutions 
where $\calN$ takes several integer values other than $0, \pm 1, \pm 2$\cite{Hata:2019dwu}.
Moreover, the author found the parameters that determine the solution is given in a closed form by using the Bernoulli numbers\cite{Hata:2019ybw}.
It is very interesting result,
however, it is difficult to understand the structure of anomaly cancellation 
due to complex combination of infinite number of terms.

In this paper, we evaluate $\calN$ and EOM in the strong sense, 
which is independent of~\cite{Hata:2019dwu,Hata:2019ybw}, 
with a pure gauge form solution represented by a finite number of terms 
constructed by $K, B, c$. 
This solutions are natural extension of the Okawa type.
We show a general formula for both of $\calN$ and EOM in the strong sense including anomaly terms.
And we discuss the possibility of anomaly cancellation of $\calN$ and EOM, simultaneously.
We concluded that we can construct an infinite number of solutions that give a $\calN$ takes an any integer value
while keeping EOM in the strong sense zero.
These solutions are different from the solutions in \cite{Hata:2019dwu,Hata:2019ybw}.
Our constructions maintain the inversion symmetry\cite{Hata:2012cy},
it means the contributions from the singularities at $K=0$
and at $K=\infty$ are equally included.

Furthermore, we mention a counterpart of the surface integration in CS theory. 
It is difficult to give a general form of ``surface integration'', since there is no concept of surface in SFT.
We discuss the evaluation of $\calN$ as a kind of surface integration for classical solutions in $KBc$ space. 
Further studies are needed in order to define surface integral that does not depend on the details of the classical solution. 

The rest of this paper is organized as follows.
In section \ref{sec:review_N=1} we review the calculation of the $\calN$ and EOM in the strong sense
of Okawa type solution by the method called $sz$-trick. 
An inevitable anomalies appear, and we explain its origin.
The aim of subsequent section is to eliminate these anomalies.
In section \ref{sec:any_N} we present the formulas of $\calN$ and EOM against solution itself for more general pure gauge type solutions in $KBc$ space. 
Although the calculations seem naively cumbersome, the results are simple.
We also check gauge invariant observable (Ellwood invariant).
In section \ref{sec:multi_brane_sol} we discuss whether $\calN$ 
can take any integer value while satisfying EOM in the strong sense.
To conclude, we discuss our results and some open problems in Section \ref{sec:discussion}.


\section{Review of Okawa type solution}
\label{sec:review_N=1}

We start by studying a class of solutions that do not depend on the details of BCFT, 
called universal,
constructed by string fields $K,B$ and $c$.
These fields satisfy following algebra under the star product\cite{Okawa:2006vm},
\begin{align}
\CR{K}{B}=0,\quad
\ACR{B}{c}=1,\quad
B^2=c^2=0,\quad
\nn\\
QK=0,\quad
QB=K,\quad
Qc=cKc.
\label{eq:KBc_algebra}
\end{align}
Okawa type solution is the following pure gauge form composed of fields $K,B,c$.
\begin{align}
\Psi&=UQU^{-1}=c\frac{KB}{G(K)}c(1-G(K)),
\label{eq:OkawaPsi}
\end{align}
where the $U$ and $U^{-1}$ are defined by
\begin{align} 
U&=1-Bc(1-G(K)),\quad U^{-1}=1+\frac{1}{G(K)}Bc(1-G(K)).
\label{eq:OkawatypeUU^-1}
\end{align}
Here, $G$ denotes an arbitrary function of $K$.
Okawa type solution has been used in the analysis of tachyon condensation 
\cite{Schnabl:2005gv,Erler:2009uj} 
and multiple brain solutions\cite{Murata:2011ep}.
Here we need to pay attention to the conditions of $G(z)$, 
which is defined by 
replacing the argument of $G(K)$ with a complex number $z$, so that 
the correlation function containing $G$ is well and uniquely defined.
We also demand the same condition as \cite{Murata:2011ep} to $G(z)$.
Although \cite{Murata:2011ep} restricts the behavior of $G$ at infinity to $G(\infty)=1$, 
we allow the condition that $G$ is a meromorphic function at infinity. 
Then $G(z)$ can be written as follows without loss of generality,\footnote{
Although $\alpha_0$ can be a pure imaginary number, we chose zero for simplicity 
because it does not change the main point of discussion below.
}
\begin{align}
G(z)&\propto \prod_{i=0}^m (z+\alpha_i)^{n_i},\quad \alpha_0=0,\quad \text{Re}[\alpha_i]>0.
\label{eq:G}
\end{align}
Hereafter we use an abbreviation $n_\infty=-n_0-\sum_{i=1}^m n_i$ for simplicity.

To calculate $\calN$ and EOM, we use the $sz$-trick introduced by Murata and Schnabl \cite{Murata:2011ep}.
This trick can be used in the calculation of correlation function including $K$, $B$ and $c$, 
reducing awful multiple integration to only two integration $z$ and $s$. 
Real number $s$ is the total width of the cylinder for which the correlation function is considered.
Since $K$ is generator of width of cylinder, $s\to\infty$($s\to 0$) corresponds to $K = 0$($K=\infty$). 
Here, we remark the $z$ integral path of $sz$-trick, which is important in the discussion below.
In the original definition of the $sz$-trick, the path of $z$ integration is defined from $-i \infty$ to $i\infty$ 
along the imaginary axis.
Since the integrand is multiplied by the factor $e^{sz}$, 
we can extend the integral path to a closed contour integral by 
adding a sufficiently large arc path in the left half plane $\operatorname{Re}z<0$, if the integrand is no worse than $O(z^{-1})$ at infinity.
Since the pole of the integrand depends on $s$, the sufficient size of the closed contour depends on $s$.
The definition of $sz$-trick does not state how to handle poles on the imaginary axis.
Unfortunately, this is an important issue regarding whether $\calN$ takes an integer value, since anomalies,  
that shift $\calN$ from integer values, arise from poles on the imaginary axis.
Bypassing the path of $z$ integration slightly to the left so as not to pick up the poles on imaginary axis is equivalent to shifting $K\to K-\eps \,\, (\eps> 0)$.
Because $K$ has a non-negative eigenvalue, this is a dangerous variable transformation that ruins inverse Laplace transformability of $G$, which is a precondition of $sz$-trick.
In this paper, we assume that an integration path include the imaginary axis to the right.
As we will see later, this assumption is provided by the $K_\eps$ regularization.
Finally, we calculate the correlation function with the following integral
\begin{align}
\frac{1}{8\pi i}\int_0^\infty\!\! ds \oint_{C_s}\frac{dz}{2\pi i}
e^{sz}\calF(z,s),
\label{eq:sz_trick}
\end{align}
where $C_s$ is a sufficiently large semicircular closed contour that includes all the poles in the left half plane,
and on the imaginary axis.
If the integrand has terms that do not vanish at infinity, 
we must treat them separately.
The detail of $\calF(z,s)$ depends on the correlation function to be calculated.  

For direct calculation of $\calN$, $\calF$ in Eq.\eqref{eq:sz_trick} is given by\cite{Hata:2012cy,Murata:2011ep}
\begin{align}
\calF(z,s)&=
-s^2\Delta(z)\frac{2z^2G'}{G}-(z\p_z-s\p_s)s^2\frac{z}{G}\Delta_s(zG).
\label{eq:calF_calN_N=1}
\end{align}
See \cite{Murata:2011ep} for the definition of $\l(\Delta_s f\r)(z)$. 
Here we omitted the terms in the form $\Delta_{2s}\l(f\circ_s g\r)=\l(\Delta_s f\r)g+f\Delta_{s}g$ 
because they do not give any contributions after $z$-integration\cite{Murata:2011ep}.
The first term of Eq.\eqref{eq:calF_calN_N=1} have no pole on the imaginary axis and
supplies the definite integer value:
\begin{align}
-\res{z} \frac{G'}{G}=-n_0-n_\infty.
\end{align}
Next, we can find the integration of the second term in Eq.\eqref{eq:calF_calN_N=1}
becomes the surface integral of $s$;
\begin{align}
\frac{1}{8\pi i}\l(\lim_{s\to \infty}-\lim_{s\to 0}\r)s^3 \res{z}e^{sz}\frac{z\Delta_s (zG)}{G},
\label{eq:anomaly}
\end{align}
In contrast to the first term, the second term  has poles on imaginary axis. 
All remains after the limit of $s$ is the contribution coming from the pole on the imaginary axis.
By combining Eqs.~\eqref{eq:calF_calN_N=1} and \eqref{eq:anomaly}, we obtain the final expression for $\calN[G]$:
\begin{align}
\calN[G]=-n_0-n_\infty +A(n_0)+A(n_\infty),
\label{eq:result_calN}
\end{align}
where
\begin{align}
A(n)=\frac{\pi^2}{3}(n^3-n)\operatorname{Re}{}_1F_1\left(2+n,4,2\pi i\right),
\label{eq:def_Anomaly}
\end{align}
where ${}_1F_1(\alpha, \beta; z)$ is the confluent hypergeometric function.
It is interesting to compare Eq.\eqref{eq:result_calN} with the most right hand side in Eq.\eqref{eq:hedgehog}.
The final expression Eq.\eqref{eq:result_calN} show that 
the value of $\calN$ is determined only by the behavior at origin and infinity of $G$,
and contributions of each term are equivalent.
Since $A(n)$ is not integer except for $n=0,\pm 1$,
$\calN$ is not an integer value generally.
However, as shown below, when $\calN$ is not integer value, the solution does not satisfy the EOM 
in the strong sense.

The reason why $\calN$ has symmetry under the exchange $n_0\leftrightarrow n_\infty$ is due to a property called the inversion symmetry of the correlation function.
There exists an map which exchange $K=0$ and $K=\infty$ while keeping
$KBc$ algebra Eq.\eqref{eq:KBc_algebra}:
\footnote{More general, following transformation keeps $KBc$ algebra 
\cite{Erler:2010zza,Masuda:2012kt,Erler:2012dz} 
$\wt{K},\wt{B}$ and $\wt{c}$ satisfy the algebra Eq.\eqref{eq:KBc_algebra}. 
$\wt{K}=g(K), \wt{B}=g(K)B/K$ and 
$\wt{c}=c(K/g(K))Bc$ for an arbitrary $g(K)$
}
\begin{align}
K\to \wt{K}=\frac{1}{K},\quad B\to \wt{B}=\frac{B}{K^2},\quad c\to \wt{c}=cK^2Bc.
\label{eq:inv_map}
\end{align}
Surprisingly, we proved that 
correlation function with an arbitrary width on a semi-infinite 
cylinder is invariant under this inversion map\cite{Hata:2012cy}.
Note that this symmetry of correlation function also holds for the regularized $K,B,c$ discussed just below. 
Under the inversion map Eq.\eqref{eq:inv_map}, 
the solution $\Psi$ of Eq.\eqref{eq:OkawaPsi} only exchanges the argument of $G(K)$ from $K$ to $1/K$.
Combining this fact, inversion symmetry of correlation function and the fact that  
$\calN$ is constructed only from $\Psi$ gives the equation $\calN[G(K)]=\calN[G(1/K)]$.

As mentioned in the introduction, $\calN$ can be rewritten as an integration form of $Q$-exact quantity,
which corresponds to the integral of the total derivative in CS theory.
Similar to the winding number$N_{CS}$, this expression clarify 
that the winding number is obtained by singularity of the solution.
The integration form of BRST-exact quantity requires parameterized solution
$\Psi_u$ which has a property interpolating between trivial vacuum at $u=0$ and non-perturbative 
configurations at $u=1$\cite{Hata:2013hba}. Then $\calN$ becomes
\begin{align}
\calN=\pi^2\int Q\l[\int_0^1\!du \, \Psi_u*\frac{d}{du}\Psi_u.\r].
\label{eq:calN=intQ}
\end{align}
However, it is generally believed the axiom that BRST-exact integral vanish holds.
Intuitively, it can be understood the integration of BRST current along the right half and the left half of string
cancel each other due to the action of $\int$.
More importantly, it is necessary for the infinitesimal gauge invariance of the action.
To be sure, however, there is a singularity at $K=0$ or $K=\infty$ in bracket of Eq.\eqref{eq:calN=intQ} 
for nonzero $\calN$.
The bracket part of Eq.\eqref{eq:calN=intQ} represent the delta functions $\delta(K)$ or $\delta(1/K)$ intuitively.
We need to introduce some appropriate regularization into Eq.\eqref{eq:calN=intQ} to extract the contributions from singularities.
To regularize the  singularity at $K=0$, we use the replacement $K\to K_\eps=K+\eps\,\,(\eps\to +0)$.
Since the eigenvalue of $K$ are non-negative, it works well.
To regularize the singularities at $K=\infty$, we use the inversion map of the $K_\eps$ regularization.
Combining both, our regularization $K_{\ee}$ is given by 
\begin{align}
K_{\ee}&=\frac{K_\eps}{1+\eta K_\eps},\quad
B_{\ee}=\frac{B}{(1+\eta K_\eps)^2},\quad
c_{\ee}=c(1+\eta K_\eps)^2 Bc,
\label{eq:K_ee}
\end{align}
where $\eps$ and $\eta$ are regularization parameter for $K=0$ and $K=\infty$, respectively. 
Note that we have to perform the replacement $K\to K_{\ee}$ after the action of the BRST charge.
This is because regularizing the bracket part in Eq.\eqref{eq:calN=intQ}, before action of the BRST charge,
corresponds to eliminate the singularities rather than regularize them.
Then RHS of Eq.\eqref{eq:calN=intQ} becomes zero for any $\Psi_u$ and contradicts on the LHS of Eq.\eqref{eq:calN=intQ}.

Applying $K_{\ee}$ regularization, Eq.\eqref{eq:calN=intQ} becomes
\begin{align}
\int \l(Q\calA[G]\r)_{K\to K_\ee}
=\int Q\calA[G(K_{\ee})]+\eps\times  \calE_{\ee}[G(K)]+\eta \times \calH_{\ee}[G(K)],
\label{eq:Qexact_N=1_reg}
\end{align} 
where $\calA$ is given by $\int\! du\,\Psi_u*d\Psi_u/du$ and 
detail expressions of $\calE_{\ee}$ and $\calH_{\ee}$ are shows in Appendix.\ref{app:calEcalH}.
The first term vanish safely, otherwise the rests can be non-trivial 
if $\calE_\ee$($\calH_\ee$) has a singularity of order $O(1/\eps)$($O(1/\eta)$). 
When we consider $K_\eps$ regularization only, the right hand side of Eq.\eqref{eq:calN=intQ} becomes $\eps \times \calE_\eps$,
where $\calE_{\eps}=\calE_{\ee}\bigl|_{\eta=0}$.
This fact indicates we can set $\eta=0$($\eps=0$) in $\calE_\ee$($\calH_\ee$), and 
we denote it as $\calE_{\eps}$($\calH_\eta$) in the following.
We can show in the Appendix\ref{app:calEcalH} that $\calH_\eta [G(K)]=\calE_\eta[G(1/K)]$ holds thanks to the inversion symmetry, 
then our evaluation reduce to 
\begin{align}
\textrm{Eq}.\eqref{eq:Qexact_N=1_reg}&=\lim_{\eps\to 0}\l(\calE_\eps[G(K)]+\calE_\eps[G(1/K)]\r).
\label{eq:N=calE+calE}
\end{align}
The result of Eq.\eqref{eq:N=calE+calE} is exactly same as the result of direct calculation of Eq.\eqref{eq:result_calN},
which indicates that $K_{\ee}$ regularization works well for the right hand side of Eq.\eqref{eq:calN=intQ}.
The $K_\eps$ regularization play a role to specify the detail of integral path along the imaginary axis in $sz$-trick, as mentioned above.
After the $K_\eps$ regularization, 
the $sz$ calculation Eq. \eqref{eq:sz_trick} of the correlation function is changed from 
$\calF(z,s)$ to $\calF (z + \eps, s)$.
If we change the variable $z_\eps\to z$ in Eq.\eqref{eq:sz_trick}, the factor $e^{-\eps s}$ is added on the integration, 
and the $z$ integral path, which was originally defined from $-i \infty$ to $i \infty$,
will bypasses the imaginary axis on the right.
As a result, by applying $K_\eps$ regularization, the $z$ integration path is determined to be $C_s$
in Eq.\eqref{eq:sz_trick}.

After $K_{\ee}$ regularization, $\Psi$ is no longer pure gauge, i.e.
$\left(UQU^{-1}\right)_{\ee}\neq U_{\ee} Q U_{\ee}^{-1}$.
This is because $K_{\ee}$ regularization breaks a part of $KBc$ algebra Eq.\eqref{eq:KBc_algebra}.
The EOM is not equal to zero and is the sum of the two terms which are of $O(\eps)$ and $O(\eta)$.
It is not clear whether the EOM is satisfied in the limit $\eps,\eta\to 0$. To check this, we should evaluate the inner product of the EOM with some variation. 
Let us consider the inner product of EOM with the solution itself :
$\calT=\int \Psi_{\ee} *\left(Q\Psi+\Psi*\Psi\right)_{\ee}$
called EOM in the strong sense. 
This quantity is necessary to link $\calN$ to the energy of the solution.
$\calT$ is not zero exactly, and
\begin{align}
\calT&=\eps \times \calE'_{\eps\eta}[G(K)]+\eta\times \calH'_{\eps\eta}[G(K)].
\label{eq:def_EOMtest}
\end{align}
See the detail of $\calE'_{\ee},\calH_{\ee}'$ in Appendix \ref{app:calEcalH}.
$\calT$ can be rewritten in the same way as Eq.\eqref{eq:N=calE+calE}
as show in Appendix.\ref{app:calEcalH},
\begin{align}
\calT=\lim_{\eps\to 0}\eps\times \l(\calE'_\eps [G(K)]+\calE'_\eps [G(1/K)]\r).
\end{align}
We can evaluate $\calE'_\eps$ by $sz$-trick, and we obtain
\begin{align}
\calT&=B(n_0)+B(n_\infty),
\label{eq:calT=B+B}
\end{align}
where $B(n)$ is given by 
\begin{align}
B(n)&=\frac{n(1+n)}{\pi}\text{Im}{}_1F_1\l(1-n,2;2\pi i\r).
\label{eq:def_B}
\end{align}
Similar to the $\calN$ anomaly $A(n)$ of Eq.\eqref{eq:anomaly},
the only poles on the imaginary axis contribute to the non-zero value of $\calE '_\eps$.
$B(n)$ does not vanish except for $n=0,\pm 1$ as well as $A(n)$.
Therefore, $\Psi$ is not a solution in the case of $A(n_0)\neq 0$.

In conclusion, $\calN$ takes an limited number of the integer values consistent with EOM.
$\calN$ and $\calT$ are determined only by the behavior of $G(K)$ at $K=0,\infty$. 
If  $|n_0|$ and $|n_\infty|$ are larger than $1$, then a pole appears on the imaginary axis, which causes anomalies 
of $\calN$ and EOM. 
Our regularization consistently defines the integral form of the BRST-exact quantity of Eq.\eqref {eq:calN=intQ}, 
while forcing us to pick up anomalies on the imaginary axis.

\section{Winding number for more general $KBc$ space}
\label{sec:any_N}

Hitherto we have considered  the simplest $U$ and its inverse Eq.\eqref{eq:OkawatypeUU^-1} in $KBc$ space.
Now, we extend $U$ and $U^{-1}$ to more general string field with ghost numbers zero in $KBc$ space as follows
\begin{align}
U&=1-\sum_{i=1}^N f_i(K) Bcg_i(K),\quad
U^{-1}=1+\frac{1}{G(K)}\sum_{i=1}^N f_i(K) Bc g_i(K),
\label{eq:UandU^-1}
\end{align}
where $G(K)$ is determined by $G=1-\sum_{i}f_i g_i$, and $g_i(K),f_i(K)$ is arbitrary function of string field $K$.
Without loss of generality, we can set the first term of $U$ equal to $1$ by using the gauge freedom. 
Then we obtain the pure gauge form $\Psi$ as
\begin{align}
\Psi&=UQU^{-1}
=\sum_{i,j=1}^N f_i(K) c\left(\delta_{ij}+\frac{a_{ij}(K)}{G(K)}\right)KBc g_j(K),
\label{eq:UQU^-1}
\end{align}
where $a_{ij}(K)$ is defined as $a_{ij}(K)=g_i(K)f_j(K)$. 
$\Psi$ with $N=1$ is equal to be Okawa type solution Eq.\eqref{eq:OkawatypeUU^-1} in the previous section.
In the following, we impose the same conditions as in the previous section on $G$ and $\l\{a_ {ij}\r\}$.
Then, similar to the expression \eqref {eq:G},
$\l\{a_ {ij}\r\}$ is also written as 
\begin{align}
a_{ij}&\propto \prod_{k=0}^m (z+\alpha^{ij}_k)^{n_k^{ij}},\quad \alpha_0^{ij}=0,\quad \operatorname{Re} \alpha_k^{ij}>0.
\end{align}
By definition, only $2N-1$ components of $\{a_ {ij}\}_{i,j=1,\cdots,N}$ are independent. 
For example, in the case of $N=4$, 
there are seven independent components, which we choose these as $a_{11},a_{12},a_{13},a_{14},a_{22},a_{33}$ and $a_{44}$, 
then the other components are written as
\begin{align}
a_{23}&=\frac{a_{22}a_{13}}{a_{12}},\quad
a_{24}=\frac{a_{22}a_{14}}{a_{12}},\quad
a_{34}=\frac{a_{33}a_{14}}{a_{13}},
\nn\\
a_{32}&=\frac{a_{33}a_{12}}{a_{13}},\quad
a_{42}=\frac{a_{44}a_{12}}{a_{14}},\quad
a_{43}=\frac{a_{44}a_{13}}{a_{14}}.
\end{align}
In the following, we employ the following components of $\{a_{ij}\}$ as independent components.
\begin{align}
\begin{pmatrix}
a_{11} & a_{12} & \cdots & a_{1N} \\
        & a_{22} &         &           \\
        &         & \ddots &        \\
       &           &         & a_{NN} \\
\end{pmatrix}
.
\end{align}
Furthermore, one of the diagonal component can be expressed by $G$, which we choose as $a_{NN}=1-G-\sum_{i=1}^{N-1}a_{ii}$.

Before going into the discussion of $\calN$, 
we mention inversion and regularization form of new $\Psi$.
Under the inversion map Eq.\eqref{eq:inv_map}, $\Psi$ changes
\begin{align}
\wt{\Psi}&=\sum_{i,j}f_i(1/K) c\left(\delta_{ij}+\frac{a_{ij}(1/K)}{G(1/K)}\right)KBcg_j(1/K).
\label{eq:Psi_inv}
\end{align}
Compare with the original $\Psi$ of \eqref{eq:UQU^-1},
the only changes are the replacement from $K$ to $1/K$ in $\l\{a_ {ij}\r\}$ and $G$. 
Thus, $\calN$ and $\calT$ are inversion symmetric quantities again, that is,
both quantity are invariant under the exchange $n_0\leftrightarrow n_\infty$ and 
$n_0^{ij}\leftrightarrow n_\infty^{ij}$. 
Therefore, it is sufficient to evaluate the contribution comes from singularities of $G$ and $\l\{a_{ij}\r\}$ at $K=0$.

We again adopt $K_{\ee}$ regularization Eq.\eqref{eq:K_ee}.
Then, the regularized new $\Psi$ is given by
\begin{align}
\Psi_{\eps\eta}&=U_{\eps\eta}QU^{-1}_{\eps\eta}
+\eps\times \sum_{i,j}f_i c\left(\delta_{ij}+\frac{a_{ij}}{G}\right)Bcg_j
+\eta\times \sum_{i,j}f_i c\left(\delta_{ij}+\frac{a_{ij}}{G}\right)K_\eps^2 Bcg_j.
\label{eq:Psi_ee}
\end{align}
Note that argument of $\l\{f_i,g_i,a_{ij}\r\}$ and $G$ are $K_{\ee}$.
$U_{\ee}$ and its inverse are defined by replacing $K$ of $\l\{f_i,g_i,a_{ij}\r\}$ and $G$ appeared in Eq.\eqref{eq:UandU^-1} by $K_{\ee}$.

\subsection{Direct calculation of $\calN$}
In this subsection, we present the formula of direct calculation of $\calN$ by $sz$ trick for any 
$\Psi$ of Eq.\eqref{eq:UQU^-1}. 
Unlike the integral form of the BRST quantity, 
the role of $K_{\ee}$ regularization in direct evaluation is not so explicit 
as long as the $z$ integration path in Eq.\eqref{eq:sz_trick} is $C_s$.
We omit the regularization for simplicity in this subsection, and we will mention about the effect of regularization 
in the end. 
Using $KBc$ algebra Eq.\eqref{eq:KBc_algebra}, we can obtain as 
\begin{align}
&\int (UQU^{-1})^3 \nonumber\\
&=
-\sum_{i,j}\VEV{a_{ji},K,\frac{K}{G}\left(1-G\right)a_{ij},K}
\nn\\
&\quad
\sum_{i,j,k}
\l\{
-\VEV{a_{ki},\frac{K}{G}a_{ij},K,\frac{K}{G}a_{jk}}
+\VEV{a_{ki},K,Ka_{ij},\frac{K}{G}a_{jk}}
+\VEV{a_{ki},\frac{K}{G}a_{ij},Ka_{jk},K}
\r\}
\nn\\
&\quad
+\sum_{i,j,k,\ell}
\VEV{a_{\ell i},\frac{K}{G}a_{ij},Ka_{jk},\frac{K}{G}a_{k\ell}},
\label{eq:calN_direct}
\end{align}
where $\langle F_1, F_2,F_3,F_4\rangle$ denote the correlation function 
$\langle BcF_1(K)cF_2(K)cF_3(K)cF_4(K)\rangle $ on a semi-infinite cylinder.
By applying the $sz$-trick formula \cite{Murata:2011ep},
\eqref{eq:calN_direct} becomes the integral Eq.\eqref {eq:sz_trick}, except that $\calF$ is replaced by
\begin{align}
\calF&=-s^2\Delta(z)\frac{2z^2G'}{G}-(z\p_z-s\p_s)s^2\frac{z}{G}\Delta(zG)
\nn\\
&\quad
-(z\p_z-s\p_s)s^2\frac{z}{G}\sum_{i,j}\left(
a_{ii}\Delta(za_{jj})
-a_{ij}\Delta(za_{ji})
\right).
\label{eq:calF_calN}
\end{align}
Here we ignore the $\Delta_{2s}$ terms, since they vanish by the integration.
The first line of \eqref{eq:calF_calN} is determined only by the diagonal components $G=1-\sum_i a_{ii}$, 
which is the same expression as Eq.\eqref{eq:calF_calN_N=1} except for the definition of $G$.
As already seen in Eq.\eqref{eq:result_calN}, it is impossible to take any integer values because of the anomaly $A(n)$.

The second line in Eq.\eqref{eq:calF_calN} is new terms appeared when $N\geq 2$.
These terms and the second term in the first line can be rewritten as the surface integration of $s$ as Eq.\eqref{eq:anomaly}.
Due to the inversion symmetry of $\calN$, it is sufficient to evaluate in the limit $s\to \infty$ corresponding to $K=0$ contribution. 
Thus let us consider the following quantity to evaluate the terms other than the first term in Eq.\eqref{eq:calF_calN}
\begin{align}
\lim_{s\to \infty}s^3\res{z}e^{sz}\frac{z}{G(z)}a_{ij}(z)\Delta(za_{k\ell}(z)).
\label{eq:test_newterm}
\end{align}
By changing the variable $z\to z/s$, we can expand integrand around $s\gg 1$.
We write the behavior of $G$ and $\l\{a_{ij}\r\}$ around $z=0$ as follows 
\begin{align}
G(z)\sim
\begin{cases}
g_0/z^{n_0} & (z\to 0)\\
g_\infty/z^{n_\infty} & (z\to \infty)\\
\end{cases}
,\quad 
a_{ij}(z)\sim 
\begin{cases}
b^{ij}_0/z^{n_0^{ij}} & (z\to 0) \\
b^{ij}_\infty/z^{n_\infty^{ij}} & (z\to \infty) \\
\end{cases}.
\label{eq:around_z=0}
\end{align}
Then we can see the $1/s$ expansion starts with $s^{-\chi}$, 
here we define $\chi$ as $\chi(i,j;k,\ell)\equiv n_0-n_0^{ij}-n_0^{k\ell}$.
If $\chi$ is positive, Eq.\eqref{eq:test_newterm} does not contribute when $s\to \infty$,
and if $\chi$ is negative, Eq.\eqref{eq:test_newterm} is not well-defined. 
A term with negative $\chi$ is not well-defined by itself, while it is allowed to be cancel in the sum of $i,j$ 
in Eq.\eqref{eq:calF_calN}.
However, we consider only $\chi\geq 0$ for simplicity from hereafter.
When $\chi(i,j;k,\ell)=0$, Eq.\eqref{eq:test_newterm} becomes 
\begin{align}
-\frac{b_0^{ij} b_0^{k\ell}}{g_0}A(n_{k\ell}),
\end{align}
where $A$ is the same as given by \eqref{eq:def_Anomaly}.
Especially in the case of $a_{ij}=1,a_{k\ell}=G$, we obtain the consistent result with the 
calculation of anomaly $A(n)$ in Eq.\eqref{eq:def_Anomaly}. 
From this calculation, we arrived at the final formula of $\calN$ for new $\Psi$;
\begin{align}
\calN&=n_0-A(n_0)-\sum_{i,j}\Theta_{ij}\frac{b^{ii}_0 b_0^{jj}}{g_0}\l(A(n^{jj}_0)-A(n^{ji}_0)\r)
\nn\\
&\quad
+\l(n_0,n_0^{ij},b_0^{ij}\leftrightarrow n_\infty, n_\infty^{ij},b_\infty^{ij}\r),
\label{eq:finalcalN}
\end{align} 
we have used $b_0^{ij}b_0^{ji}=b_0^{ii}b_0^{jj}$ from the definition of $a_ {ij}$.
$\chi(i,j)$ is defined as 
\begin{align}
\chi(i,j)=\chi(i,j;j,i)=n_0-n_0^{ij}-n_0^{ji}=n_0-n_0^{ii}-n_0^{jj}.
\label{eq:chi}
\end{align}
and $\Theta_{ij}$ is defined as
\begin{align}
\Theta_{ij}&=
\begin{cases}
1 & \chi(i,j)=0, \\
0 & \chi(i,j)>0, \\
\end{cases}
\end{align}
and $n_\infty^{ij}$ is defined as $-n_0^{ij}-\sum_{k=1}^M n_k n_k^{ij}$.
The expression Eq.\eqref{eq:finalcalN} includes degree of freedom that are not independent.
From the above result, we will construct the concrete expression for $\calN$ in Sec.\ref{sec:multi_brane_sol}. 

Before ending this subsection, we mention on the direct calculation of $\calN$ with regularization.
If we apply $K_{\eps}$ regularization to Eq.\eqref {eq:calN}, 
the calculation of the first term does not have essential change, 
but the remaining terms changes as
\begin{align}
\eps \times \int_0^\infty\!\! ds\,\,e^{-\eps s}\,s^3  f(s)
+\l(\lim_{s\to \infty}-\lim_{s\to 0}\r)e^{-\eps s}s^3\cdots,
\label{eq:anomaly_offdiagonal}
\end{align}
where
\begin{align}
f(s)&=\frac{1}{8\pi i}\oint_{C_s}\frac{dz}{2\pi i}
e^{sz}\frac{z}{G}\left(
\sum_{i}\sum_{j\neq i}a_{ii}\Delta(za_{jj})
-\sum_{i\neq j}a_{ij}\Delta(za_{ji})
\right).
\end{align}
Thanks to the suppression factor $e^{-\eps s}$, 
the surface term in Eq.\eqref{eq:anomaly_offdiagonal} vanish 
\footnote{
If we regularize $K=\infty$, we can see that the limit of $s\to 0$ also vanish.
}.
Taking $\eps$ to be zero in Eq.\eqref{eq:anomaly_offdiagonal}
is equivalent to picking up the behavior of $s^3 f(s)$ at $s=\infty$,
as long as $\lim_{s\to \infty}s^3 f(s)$ is well defined.
Therefore,we obtain the same result as the direct calculation of $\calN$.

\subsection{Integration form of BRST-exact quantity and surface integration}
In this subsection, we give an evaluation of $calN$ as the integration form of the BRST-exact quantity.
The outline of the calculation is the same as that of the Okawa type.
In addition, we mention about ``the surface integration'' of CSFT. 
Although this is a specific argument expressed in terms of $sz$-trick, 
the discussion widely holds for the integral form of BRST-exact quantities.

In order to investigate the integration of BRST-exact quantity for the $\Psi$,
as in Okawa solution,
we introduce the parameterized $\Psi_u$ defined as follows.
\begin{align}
\Psi_u&=u\sum_{i,j}f_i(K)c\left(\delta_{ij}+u\frac{a_{ij}(K)}{G_u(K)}\right)KBcg_j(K),
\end{align}
where $G_u$ is defined as $G_u=1-u\sum_{i} a_{ii}$.
Then we obtain the integration of the $Q$-exact form of $\calN$ as
\begin{align}
\calN/\pi^2
&=\int Q\l(\sum_{i,j,k}\int_0^1\!du\,
cK\CR{\frac{ua_{ij}}{G_u}}{c}\frac{ua_{jk}K}{G_u}Bca_{ki}
\r).
\label{eq:intQexact}
\end{align}
By applying the $K_{\ee}$ regularization, we obtain the following form
\begin{equation}
  \begin{split}
  \calN/\pi^2 =\eps\times \calE_{\eps\eta}[G(K),a_{ij}(K)]+\eta \times\calH_{\eps\eta}[G(K),a_{ij}(K)]\Bigr).
  \end{split}
  \label{eq:calE_calH}
\end{equation}
Here we have omitted the term $\int Q\l[(\Psi_u*d\Psi_u/du)_\text{reg}\r]$ because it is safely zero.
The details of $\calE_{\ee}$ and $\calH_{\ee}$ are shown in Appendix.\ref{app:calEcalH}.
By inversion symmetry of correlation function, we can show 
\begin{align}
\calH_{\eta}[G(K),a_{ij}(K)]=\calE_{\eta}[G(1/K),a_{ij}(1/K)],
\end{align}
where we denote $\calH_{\eta}$ as $\calH_{\ee}$ with $\eps$ set to zero and $\calE_{\eps}$ as $\calE_{\ee}$ with $\eta$ set to zero,
and $\calE_{\eta}$ is obtained by replacing $\eps$ with $\eta$ in $\calE_{\eps}$.
Therefore $\calN$ is reduce to be the expression including only $\calE_\eps$;
\begin{align}
\calN&=\pi^2 \lim_{\eps\to 0}\eps\times \l(\calE_\eps[G(K),a_{ij}(K)]+\calE_{\eps}[G(1/K),a_{ij}(1/K)]\r).
\end{align}
The $\calE_\eps$ is same as the integral Eq.\eqref {eq:sz_trick}, except that $\calF$ is replaced by
\begin{align}
\calF&=\l(
-2\Delta \l(\frac{z}{G_u}\r) \frac{zG}{G_u}
-\Delta(zG)\frac{z}{G_u^2}
+\Delta (z)\frac{zG}{G_u^2}
\right.\nn\\
&\left.\quad
+\frac{u(1+G_u)}{G_u^2}\sum_{i\neq j}\l(
\Delta(z a_{ij})za_{ji}
-
\Delta(z a_{ii})za_{jj}
\r)
\r)',
\label{eq:calF_Qexcat}
\end{align}
where $'$ denote the differentiation with respect to $z$. 
The first line is given by only diagonal component $G=1-\sum_i a_{ii}$ 
and the same expression as in the case of $N=1$. 
It has already been proven in \cite{Hata:2012cy} to be equivalent to the result of 
the direct $\calN$ calculation Eq.\eqref{eq:result_calN}. 
In the second line, using following equation
\begin{align}
\int_0^1\! du\, \frac{u(1+G_u)}{G_u^2}=\frac{1}{(1-G)^2}\l[G_u+\frac{1}{G_u}\r]^{u=1}_{u=0}
=\frac{1}{G},
\end{align}
we notice that it has the same expression as the new anomaly term in Eq.\eqref{eq:calF_calN}.
Therefore, we can reproduce the result of direct calculation of $\calN$ 
from the integral of BRST-exact quantity with $K_{\ee}$ regularization for the new $\Psi$.
Note that if we performed the $u$ integration after the $z$ or $s$ integration, 
we numerically checked that is equal to the result of the direct calculation.

Now, recall the fact that the winding number $N_{CS}$ can be written in the surface integral, 
as we saw in Eq. \eqref{eq:hedgehog}.
We already know that the result of the $\calN$ calculation Eq.\eqref{eq:result_calN}
resembles the most right hand side of 
Eq. \eqref {eq:hedgehog}. 
What can be inferred from this result is that $\calN$ can be expressed as a surface integral over $K$ or $s$. 
This is because $s$ is the total width of the cylinder and $K$ is the generator of width. 
In fact, we can prove that the correlation function $\int Q(BcF_1(K)cF_2(K)cF_3(K))$ 
is expressed by the surface integral of $s$ in $sz$-trick for arbitrary functions $F_1(K),F_2(K)$ and $F_3(K)$
as seen in Eq.\eqref{eq:Qexact_szformula}.
From the $sz$-trick formula, we arrive at
\begin{align}
\int Q\l(BcF_1cF_2cF_3\r)&=\frac{1}{8\pi^3 i}\int_0^\infty \!ds\,
\int_{-i\infty}^{i\infty}\frac{dz}{2\pi i}e^{sz}\calF(z,s),
\label{eq:Qexact_sz}
\end{align}
where $\calF$ is given by
\begin{align}
\calF&=\calG\l(F_1,F_2,F_3,K\r)+\calG\l(F_1,K,F_2,F_3\r)-\calG\l(F_1,F_2,K,F_3\r)
\nn\\
&=\l(z\p_z-s\p_s \r)s^2\l(\Delta_{2s}(F_1\circ_s F_2)F_3-\Delta_s(F_1F_2)F_3\r).
\label{eq:calF_Qexact_v1}
\end{align}
Especially, if $F_2=K$, it becomes 
$
\calF=2\pi i(z\p_z-s\p_s)(sF_3\Delta_{2s}^2F_1),
$
this is just the same expression in \cite{Murata:2011ep}.
If we can add an infinite arc to integration path on the left half plane in order to make a closed contour, 
Eq. \eqref{eq:Qexact_sz} becomes the surface integration of $s$
\begin{align}
&\int Q(BcF_1cF_2cF_3)
\nn\\
&=
\frac{1}{8\pi^3 i}\l(\lim_{s\to \infty}-\lim_{s\to 0}\r)s^3
\oint_{C_s}\frac{dz}{2\pi i}
e^{sz}
\l(
\l(\Delta_s(F_1F_2)-\Delta_{2s}(F_1\circ_s F_2)\r)F_3
\r).
\label{eq:Qexact_szformula}
\end{align}
If closed contour does not include any poles on the imaginary axis, 
$e^{sz}$ act as a suppression term in the limit of $s\to\infty$.
The contribution from $s\to 0$ will be zero 
due to the factor of $s^3$ when $F_1, F_2, F_3$ satisfy the appropriate conditions.
Therefore, it seems that the axiom that the BRST-exact integration vanish is valid naively.

However, we encounter the different situation when $z$-integral picks up a pole on the imaginary axis.
As already said, $z$ integration path we adopt is a closed contour $C_s$ 
that includes the imaginary axis.
To be sure, we did not evaluate the value of $\calN$ directly from the form of $\int Q\calA$ above, 
but evaluated by applying $K_\eps$ regularization to break BRST-exactness.
However, the form of $\int Q \calA$ can produces the same result as 
that obtained by $K_\eps$ regularization.
To explain this,
we apply $K_\eps$ regularization in Eq.\eqref {eq:Qexact_sz}, it becomes 
\begin{align}
&\eps \times \int\!  cF_1(K_\eps)cF_2(K_\eps)cF_3(K_\eps)
\nn\\
&=
\eps\times \frac{1}{8\pi^3 i}\int_0^\infty \!\!ds\, e^{-\eps s}s^2
\res{z}\!e^{sz}\p_z\l\{\l(\Delta_{2s}(F_1\circ F_2)-\Delta_s(F_1F_2)\r)F_3\r\}.
\label{eq:intQcalA_reg}
\end{align}
Here, we have omitted the $\int Q\l(Bc F_1(K_\eps)c F_2(K_\eps)cF_3(K_\eps)\r)$ because 
there does not exist any poles on the imaginary axis in the $sz$-trick.
After partial integration, Eq.\eqref{eq:intQcalA_reg} with $\eps$ set to zero is same as $s\to \infty$
term in Eq.\eqref{eq:Qexact_szformula}, unless the $s$ integral diverges.
Therefore, we can say that integration of BRST exact quantity \eqref{eq:Qexact_szformula} can take a non-zero value.
Note that we discussed only $\lim_{s\to \infty}$ term in Eq.\eqref{eq:Qexact_szformula} 
which is necessary for following discussion.

Now, let us return to $\calN$ of Eq.\eqref{eq:intQexact}.
This becomes the surface integration of $s$ as discussed above,
\begin{align}
\calN&=\frac{1}{8\pi i}\l(\lim_{s\to \infty}-\lim_{s\to 0}\r)\int_0^1\!\! du \oint_{C_s}\frac{dz}{2\pi i}e^{sz}\calF(z,s,u),
\label{eq:QexactcalN_sz} 
\end{align}
where
\begin{align}
\calF(z,s,u)&=\sum_{i,j,k}\l\{
\Delta_s(a_{ki}\frac{uza_{ij}}{G_u})-\Delta_{2s}\l(a_{ki}\circ_s \frac{uza_{ij}}{G_u}\r)
\r\}\frac{uza_{jk}}{G_u}
\nn\\
&\quad-\sum_{i,j}
\l\{
\Delta_s(za_{ji})-\Delta_{2s}(a_{ji}\circ_s z)
\r\}\frac{1-G_u}{G_u^2}uza_{ij}.
\label{eq:calF_calN=intQcalA}
\end{align}
The poles of Eq.\eqref{eq:calF_calN=intQcalA} exist only on the left half plane 
$\operatorname{Re}z<0$, except for $u=0$.
If we perform the integration with respect to $u$ after limiting $s$ to zero or infinity, Eq.\eqref{eq:QexactcalN_sz} becomes zero in both cases.
However, if the $u$ integration is performed before the limit of $s$, 
Eq.\eqref{eq:QexactcalN_sz} gives the same result as direct calculation of $\calN$.

Certainly, we could rewrite Eq. \eqref {eq:intQexact} into the surface integration form.
Although the integration form of BRST-exact quantity is formally given regardless of the details of the solution, 
the surface integration form depends on the property of the solution.
Furthermore, the correlation function can be written with $sz$-trick explicitly.
In these respects, the above discussion  did not give a general solution.
Even so, since the statement holds for any $F_1,F_2,F_3$, it will give suggestions to the general definition of surface integration in CSFT.

\subsection{EOM against solution itself}
It was found that the value of winding number is provided by the singularity of $\Psi$. 
Regularization is necessary to safely handle singularities and extract them properly.
In the previous section, we saw that $K_{\ee}$ regularization works well for 
the integration form of the BRST-exact quantity, 
however, it should be noted that this regularization breaks the equation of motion at $O(\eps)$.
Moreover, the integration path specified by $K_\eps$ regularization picks up anomalies derived from poles on the imaginary axis, and thus $\calN$ does not takes an integers for general $\Psi$.

The equation of motion for $\Psi_{\ee}$ in Eq.\eqref{eq:Psi_ee} is
\begin{align}
&Q\Psi_{\eps\eta}+\Psi_{\eps\eta}*\Psi_{\eps\eta}\nn\\
&=\eps\times
\sum_{i,j}f_i c\frac{K_\eps^2}{K_{\ee}}
\left(\delta_{ij}+\frac{a_{ij}}{G}\right)cg_j
+\eta\times \sum_{i,j}f_i c K_\eps^2
\CR{\frac{1}{K_{\eps\eta}}\left(\delta_{ij}+\frac{a_{ij}}{G}\right)}{c}
K_\eps^2Bcg_j.
\label{eq:EOManomaly}
\end{align}
We extend straightforwardly the discussion of $\calT$ Eq.\eqref{eq:def_EOMtest} as in the Okawa type solution
\begin{align}
\calT
&=\eps\times \calE'_{\eps\eta}[G(K),a_{ij}(K)]
+\eta\times \calF'_{\eps\eta}[G(K),a_{ij}(K)].
\label{eq:calE'_calH'}
\end{align}
Details of $\calE'_{\ee}$ and $\calF'_{\ee}$ are given in Appendix.\ref{app:calEcalH}. 
As shown in Eq.\eqref{eq:def_EOMtest}, this $\calT$ also can be expressed by $\calE'_\eps$ alone thanks to inversion symmetry.
Defining $\calE'_{\eps}$ and $\calH'_\eta$ as $\calE'_{\eps,\eta=0}$ and $\calH'_{\eps=0,\eta}$, respectively, gives
\begin{align}
\calT&=\lim_{\eps\to 0}\eps\times
\left(
\calE'_{\eps}[G(K),a_{ij}(K)]
+\calE'_{\eps}[G(1/K),a_{ij}(1/K)]
\right).
\label{eq:calT_invsymmetricform}
\end{align}
The $\calF$ in Eq.\eqref{eq:sz_trick} for $\calE'_\eps[G(K),a_{ij}(K)]$ becomes
\begin{align}
\calF&=
-4s^2\Delta_s(z)\frac{zG'}{G}+2(z\p_z-s\p_s)s^2 G\Delta\l(\frac{z}{G}\r)
+s^2\l(zG\Delta\frac{z}{G}\r)'
\nn\\
&\quad
+\sum_{i\neq j}\l[
(z\p_z-s \p_s)s^2a_{ii}\Delta_s\frac{za_{jj}}{G}+
s^2\l(za_{ii}\Delta_s\frac{za_{jj}}{G}\r)'
\right.\nn\\
&\left.\quad
-
\l\{
(z\p_z-s \p_s)s^2a_{ij}\Delta_s\frac{za_{ji}}{G}+
s^2\l(za_{ij}\Delta_s\frac{za_{ji}}{G}\r)'
\r\}
\r].
\label{eq:calT_calF}
\end{align}
The first line leads to the same result as Eq.\eqref{eq:calT=B+B}, 
and the remaining terms are given as follows using Eq.\eqref {eq:around_z=0};
\begin{align}
\calT&=B(n_0)+\sum_{i\neq j}\Theta_{ij}\frac{b^{ij}_0 b^{ji}_0}{g_0}\l(
B(n_0^{jj})-B(n^{ji}_0)
\r)
\nn\\
&\quad
+\l(n_0,n_0^{ij},g_0,b_0^{ij}\leftrightarrow n_\infty,n_\infty^{ij},g_\infty,b_\infty^{ij}\r).
\label{eq:finalcalT}
\end{align}
Replacing $A$ with $B$,
this expression is the same as the anomaly term in $calN$ in Eq.\eqref{eq:finalcalN}.

\subsection{Gauge invariant observables}
In this section, we discuss about the gauge invariant observable(GIO)
\cite{Hashimoto:2001sm,Gaiotto:2001ji,Ellwood:2008jh},
\begin{align}
\VEV{I|c(i)c(-i)V(i,-i)|\Psi},
\label{eq:def_GIO}
\end{align}
where $V$ is an on-shell closed string vertex operator,
$\bra{I}$ is an identity string field.
GIO is an important tool for extracting BCFT information from classical solutions.
We assume $\calI (y)$ is defined as follows, 
\begin{align}
\calI(y):=2\pi^2 \int c \bar{c}V(iy,-iy) \Psi,
\label{eq:calI}
\end{align}
where it is written in sliver coordinates.
Then GIO is given as $\lim_{y\to \infty}\calI(y)$.
Calculating GIO in the same way as \cite{Murata:2011ep} gives
\begin{align}
\lim_{y\to \infty}\calI(y)&=
\calA_0^\text{disk}(V^m)\lim_{z\to 0}\left(
-zG'+\sum_{i,j}\frac{z}{G}a_{ij}a'_{ji}
\right)
=-\calA_0^\text{disk}(V^m)\lim_{z\to 0}\frac{zG'}{G}
\nn\\
&=n_0\times \calA_0^\text{disk}(V^m).
\label{eq:EI_general}
\end{align} 
This result indicates $\Psi$ represents $n_0$ copies of the original D-brane. 
Unlike $\calN$, GIO has no anomaly.
 
It should be noted that GIO does not equally include the contributions from $K=0$ and $K=\infty$. 
Unfortunately, since GIO contains matter operator $V$, it cannot be shown that GIO 
is an inversion symmetric quantity.
We comment on a inversion symmetric GIO in Sec.\ref {sec:discussion}.

\section{Construction of winding number for arbitrary integer values}
\label{sec:multi_brane_sol}

In this section, we construct a solution where $\calN$ takes an integer value 
while keeping EOM in the strong sense $\calT$ to zero.
We should consider both contributions from the singularities at $K=0$ and $K=\infty$, but for simplicity, we discuss the contribution only from $K = 0$ in this section.
That is, we investigate a combination of $G$ and $\l\{a_{ij}\r\}$ where both $\calN$ anomaly,
\begin{align}
A(n_0)+\sum_{i,j}\Theta_{ij}\frac{b_0^{ij}b_0^{ji}}{g_0}\l(A(n_0^{jj})-A(n_0^{ji})\r),
\label{eq:totanomaly}
\end{align}
in Eq.\eqref{eq:finalcalN} and EOM anomaly, $\calT$ in Eq.\eqref{eq:finalcalT}, are zero.
In the previous section, we showed that $\calN$ and $\calT$ could diverge unless $\Psi$ is properly adjusted,
which feature did not exist when $N=1$.
We have to take care that the index $\chi(i,j)$ Eq.\eqref{eq:chi} is a non-negative integer so that $\calN$ and $\calT$ are well defined.
This conditions can reduces the degree of freedom prepared at the beginning, but if $N$ is chosen to be 3 or more, there is still enough freedom to make both $\calN$ anomaly and $\calT$ to zero as seen below.

\subsection{$N=2$ case}
Let us start with the simplest case, i.e. $N = 2$.
To achieve $\chi_{12}=\chi_{21}=0$ under the constraint $G=1-a_{11}-a_{22}$, 
either $n_0^{11}$ or $n_0^{22}$ should be $n_0$ and the other should be $0$.
Since the roles of $a_{11}$ and $a_{22} $ are symmetric, let us consider the case of 
$n_0^{11}=n_0, n_0^{22}=0$ from here after.
In addition, it is hold that $b_0^{11}=-g_0$ for $n_0>0$, and $b_0^{22}=1$ for $n_0<0$.
Therefore Eq.\eqref{eq:totanomaly} and Eq.\eqref{eq:finalcalT} becomes
\begin{align}
&
A(n_0)+\gamma\l(A(n_0)-A(n_0^{12})-A(n_0-n_0^{12})\r),
\label{eq:anomaly_N=2}
\\
&
B(n_0)+\gamma\l(B(n_0)-B(n_0^{12})-B(n_0-n_0^{12})\r),
\label{eq:EOM_anomaly}
\end{align}
where coefficient $\gamma$ is define as $\gamma=-b_0^{22}$ for $n>0$ and $\gamma=b_0^{11}/g_0$ for $n<0$.
Thus, for an integer $n_0$, our task is finding $n_0^{12} \in \mathbb{Z}$ and $\gamma$ 
so that the two anomalies Eq.\eqref{eq:anomaly_N=2} and Eq.\eqref{eq:EOM_anomaly}
to be zero at the same time.The results are
\begin{align}
n_0&=2\quad
\left(n_{12},\gamma\right)=\l(-1,1/3\r),\,\,
\l(1,-1\r),\,\,
\l(3,1/3\r),
\nn\\
n_0&=3\quad
\left(n_{12},\gamma\right)=(1,-3/4),\,\,(2,-3/4),
\nn\\
n_0&=-2\quad
\left(n_{12},\gamma\right)=(-1,-1).
\end{align}
Here, we omitted solutions satisfying $\calN=\pm 1$ that can be constructed with $N=1$.

\subsection{$N=3$ case}
In the case of $N=3$, we separately describe the case where $n_0$ is positive and negative.

At first, we consider the case that $n_0$ is positive.
From the condition $1-G=\sum_i a_{ii}$,
it is impossible that all $n_0^{11}, n_0^{22},n_0^{33}$ are smaller than $n_0$, and 
at least one of $n_0^{11}, n_0^{22}, n_0^{33}$ should be equal to or grater than $n_0$.
Therefore all three $\chi(1,2), \chi(1,3), \chi(2,3)$ cannot be zero at the same time.
Here we set two $\chi(1,2),\chi(1,3)$ to zero and set $\chi(2,3)> 0$. 
To realize this, take for example $n_0^{11}=n_0$ and $n_0^{22}=n_0^{33}=0$.
In this case, $b_0^{11}=-g_0$ holds.

Next, consider the case $n_0<0$. 
At least one of $n_0^{11}, n_0^{22}, n_0^{33}$ will be zero by condition $G =1-\sum_i a_{ii}$
while keeping $\chi\geq 0$.
Again, all three $\chi(1,2),\chi(1,3),\chi(2,3)$ cannot be zero at the same time, 
then we consider $\chi_{12}=\chi_{13}=0$ and $\chi_{23}>0$ case.
In this case, we can take $n_0^{11}=0, n_0^{12}=n_0^{13}=n_0$ and $b_0^{11}=1$.

As the result, the both anomalies in Eq.\eqref{eq:totanomaly} are reduce to 
\begin{align}
A(n_0)
&
+\gamma^2\left(A(n_0)-A(n_0^{12})-A(n_0-n_0^{12})\right)
\nn\\
&
+\gamma^3\left(A(n_0)-A(n_0^{13})-A(n_0-n_0^{13})\right),
\label{eq:anomaly_N=3}
\end{align}
where $\gamma^{i}=-b_0^{ii}$ for $n_0>0$ and $\gamma^i=b_0^{ii}/g_0$ ($i=2,3$).
$\calT$ can be obtained by simply replacing $A$ with $B$ in the above expression.  
For any integer values of $n_0$, there exist appropriate two coefficients $\gamma^2,\gamma^3$ to vanish 
both anomalies.

\subsection{$N\geq 4$ case}
It is straightforward to generalize the above discussion to $N\geq 4$.
From the condition $1-G=\sum_{i}a_{ii}$, at least one of the elements, $n_0^{11},n_0^{22},\cdots ,n_0^{NN}$, should be equal to or grater than $n_0$.
If there exist elements greater than $n_0$, at least two elements are required in order to satisfy the condition $1-G=\sum_i a_{ii}$.
This is not an appropriate situation because the combination of these elements produces negative $\chi$.
Therefore, only one of $n_0^{11},n_0^{22},\cdots ,n_0^{NN}$ should be equal to $n_0$, 
here we choose $n_0^{11}$, and the remaining elements should satisfy $n_0^{ii}\leq 0$ from the condition $\chi^{1i}\geq 0$.
For $i$ that satisfy $n_0^{ii}=0$, the additional anomaly term $A(n_0)-A(n_0^{1i})-A(n_0-n_0^{1i})$ to Eq.\eqref{eq:totanomaly} is required by the condition $\chi(1,i)=0$.
For $i$ that satisfy $n_0^{ii}<0$, additional anomaly element never appear.
In addition, the anomaly term $A(n_0^{ii})+A(n_0^{jj})-A(n_0^{ij})-A(n_0^{ji})$ does not appear either, because $\chi(i,j)$ is positive for all $i,j=2,3,\cdots,N$. 
Thus, if we define
\begin{align}
I_{>}=\l\{ i | n_0^{ii}=0 ,2\leq i \leq N\r\},
\end{align}
our task is to find $n_0^{1i} \in \mathbf{Z}$ and $b_0^{ii}$ which satisfy the following two equations;
\begin{align}
&A(n_0)-\sum_{i\in I_{>}} b_0^{ii}\l(A(n_0)-A(n_0^{1i})-A(n_0-n_0^{1i})\r)=0,
\nn\\
&B(n_0)-\sum_{i\in I_{>}} b_0^{ii}\l(B(n_0)-B(n_0^{1i})-B(n_0-n_0^{1i})\r)=0.
\end{align}
For the case of $n_0<0$, from the same discussion as above, one element in $n_0^{11},n_0^{22},\cdots,n_0^{NN}$ should be zero and is chosen as $n_0^{11}$.
Then if we define
\begin{align}
I_{<}=\l\{ i | n_0^{ii}=n_0 ,2\leq i \leq N\r\},
\end{align}
our task is to find $n_0^{1i} \in \mathbf{Z}$ and $b_0^{ii}/g_0$ which satisfy the following two equations;
\begin{align}
&A(n_0)+\sum_{i\in I_{<}} \frac{b_0^{ii}}{g_0}\l(A(n_0)-A(n_0^{1i})-A(n_0-n_0^{1i})\r),
\nn\\
&B(n_0)+\sum_{i\in I_{<}} \frac{b_0^{ii}}{g_0}\l(B(n_0)-B(n_0^{1i})-B(n_0-n_0^{1i})\r).
\end{align}
As a result, we conclude that there is enough degrees of freedom for any $n_0$ to satisfy the two conditions if $N\ge3$.
Moreover, in this case, we find infinite number of solutions that satisfy $\calN=n_0$, and we need to impose the appropriate EOM conditions to these solutions.

\section{Summary and discussion}
\label{sec:discussion}

In this work, we have studied whether $\calN$ of Eq.\eqref{eq:calN} can take an integer values keeping EOM to zero in the strong sense $\calT$. 
This work is inspired by the similarity of the algebraic structure of the CSFT and CS theory, which suggest that $\calN$ is a winding number.
For Okawa type solution, we have already discussed this problem in our previous paper \cite{Hata:2011ke,Hata:2012cy}.
Unfortunately, $\calN$ takes only some integer values due to some kind of anomalies.
Therefore, we investigated in more general $KBc$ space than Okawa solution, by extending $U$ as Eq.\eqref{eq:UandU^-1} in this paper.
We have presented the general formula of $\calN$, EOM against solution itself and the gauge invariant observable.
Our results showed that at least three terms in $U$ and $U^{-1}$ are sufficient to eliminate 
the two kind of anomalies, i.e. $A(n)$ in $\calN$ and $B(n)$ in $\calT$.
We also discussed about the surface integration form of $\calN$.
We demonstrated that a general integration form of BRST quantity consisting of $K,B$ and $c$ 
can take non-zero values by direct calculation in $sz$-trick. 
Moreover, it can be written in the surface integration of the $s$ in $sz$-trick,
where $s$ denote total width of cylinder for which we are consider the correlation function.

In the paper \cite{Mertes:2016vos}, $LKBc$ algebra, which is an natural extension of $KBc$ algebra, 
was introduced. 
As pointed out in \cite{Mertes:2016vos,Zeze:2019bbe}, 
$L$ can be understood as a derivative of $K$. 
It might be possible to rewrite the surface integration form of $\calN$ to a more universal form by using $L$.

We gave the formula for $\calN$ and $\calT$ only if each summrand in Eq.\eqref{eq:calN_direct} and Eq.\eqref{eq:calT_calF} satisfies the condition $\chi(i, j) \geq 0$. 
if $\chi(i,j) < 0$, it is difficult to give the explicit solution, because each summrand in Eq.\eqref{eq:calN_direct} and Eq.\eqref{eq:calT_calF} diverge.
There is a slight possibility to cancel the divergence in the sum and give different solutions from this paper.
However, this cancellation procedure would require the fine tune of parameters in $\calN$ and $\calT$ simultaneously.
Also, since the formula of Eq. \eqref {eq:test_newterm} in the case of $\chi (i, j)< 0 $ 
depends on the higher order derivative of $G$ and $\l\{a_{ij}\r\}$, 
it is hard to say that $\calN$ and $\calT$ are determined only by the behavior of the solution at $K=0$ and 
$K=\infty$.

In this paper, we did not mention about the problem of which variation EOM vanish.
We only examined the inner product of EOM and the solution itself. 
This is not enough to determine the solution, and we should check the EOM for more various conditions.
It should be mentioned here about the EOM against a state in the Fock space.
For Okawa solution $\Psi = c\frac{KB}{G}c(1-G)$,
\cite{Murata:2011ep} gave the coefficient in front of $c_1c_0\ket{0}$ in EOM by $C(n_0)$, where $C(n)$ is defined as
\begin{equation}
C(n)=\frac{\pi n(n+1)}{4} \operatorname{Re}_1 F_1(2+n,3,2\pi i).
\end{equation}
$C(n)$ vanish only for $n=-1,0$.
In contrast, if we calculate for $\Psi = fc\frac{KB}{G}c g\ \ (G=1-fg)$ instead of the above $\Psi$, 
the result is not $C(n_0)$. To make matter worse, it generally diverges when $\eps \to 0$.
If the behavior of $f(z)$ near $z=0$ is $z^{-n_0^f}$, then the leading term when expanded in $\eps$ is 
\begin{equation}
  \frac{1}{\eps}\left( C'(n_0) - C'(n_0-n_0^f) -C'(n_0^f) \right) ,
\end{equation}
where 
\begin{equation}
C'(n)=\frac{ n(n+1)}{4} \operatorname{Im}_1 F_1(2+n,3,2\pi i).
\end{equation}
When $n_0^f=0$ or $n_0^f=n_0$, the term proportional to $1/\eps$ vanish, and the next order matches $C(n_0)$.
This calculation can be extended to general $\Psi$ in Eq.\eqref{eq:UQU^-1}.
However, similar situation generally occur even if $\Psi$ satisfies the conditions, which is necessary for $\calN\in \mathbf{Z}$ and $\calT=0$.
What to keep in mind is that we do not have any clear reason to choose Fock space as the variation, since Fock state is not a natural state on non-perturbative vacuum.
At present, it has not been possible to judge the above result correctly yet.
This is an important problem that we should tackle in the near future.

This paper also did not cover the relation between our result and the solution in \cite{Hata:2019dwu}.
In \cite{Hata:2019dwu}, in addition to $\calT=0$, reality conditions are imposed on the solution.
We need to consider the reality conditions to construct a physical classical solution.

In contrast to $\calN$, GIO is determined only by $G (K)$, and has no anomaly for any solution. 
And GIO can be obtained from $O(\eps)$ term in Eq.\eqref{eq:Psi_ee} when the regularization is applied.
This property is the same as $\calN$.
Furthermore \cite{Baba:2012cs} showed that GIO, for a certain case with matter operator $V=\p X^0 \bar{\p} X^0$, is equal to $\calN$.
This results indicate that the GIO is also topological quantity.
Based on the proof of \cite{Baba:2012cs}, we stated in the previous paper\cite{Hata:2013hba} 
that inversion symmetric GIO for $V=\p X^0 \bar{\p} X^0$ is defined as 
\begin{align}
\l(\lim_{y\to \infty}-\lim_{y \to 0} \r)\calI(y),
\label{eq:invGIO}
\end{align}
where $\calI(y)$ is defined as Eq.\eqref{eq:calI}.
This expression is different from  the ordinary GIO $\lim_{y\to \infty}\calI(y)$ in Eq.\eqref{eq:def_GIO}.
Since GIO contains matter operator $V$, we cannot show Eq.\eqref{eq:invGIO} has the inversion symmetry 
directly.
However, we confirmed 
for some multi brane solution that 
$\lim_{y\to \infty}\calI(y)$ and $\lim_{y\to 0}\calI(y)$  picks the contribution form the singularity at $K=0$ and $K=\infty$, respectively.
Also, Eq.\eqref{eq:invGIO} seems the surface integration of $y$.
With these reasons in mind, we conjecture that Eq.\eqref{eq:invGIO} is genuine gauge invariant observable for any matter operator $V$.
It is interesting future problem.

Finally, we will make a brief comment on a solution in non-universal class\cite{Erler:2014eqa} 
from the view of the winding number. Here we call it as EM solution.
Although the EM solution is based on the tachyon vacuum solution ($\calN =-1$), 
we can formally construct the solution based on the multi-brane solution $\Psi_{cl}=cKB/G(K)c(1-G(K))$ 
in the same way as EM solution by using the fact $Q_{cl}(BG(K))/K=1$ hold algebraically.
Here $Q_{cl}$ is the BRST charge around the solution $\Psi_{cl}$.
According to \cite{Erler:2014eqa}, the energy of this solution is 
the product of VEV of boundary condition changing operator and the winding number.
Therefore, even in the EM solution, the search for topological structures in the $KBc$ space is still important.

\section*{Acknowledgments}

We would like to thank H.Hata and M.Schnabl for useful discussions at various stages of this work.
We thank the organizers of the 2019 workshop ``String Field Theory and String Perturbation Theory'' in Florence, 
for providing a stimulating environment during part of this research. 

\newpage
\appendix

\section{Details of calculation about $\calN$ and $\calT$}
\label{app:calEcalH}

In this appendix, we present the details of $\calE_{\ee},\calH_{\ee}$ in Eq.\eqref{eq:calE_calH}
and $\calE'_{\ee},\calH'_{\ee}$ in Eq.\eqref{eq:calE'_calH'}.
And we demonstrate that $\calE$ and $\calH$ ($\calE'$ and $\calH'$) are non trivially 
connected, thanks to the inversion symmetry of the correlation function.

First, we give a concrete expression of $\calE_{\ee}$ and $\calH_{\ee}$
\begin{align}
\calE_{\eps\eta}[G(K)]&=
\sum_{i,j,k}\int_0^1\!\! du\l[
\int\! ca_{ki}c_{\eps\eta}\frac{ua_{ij}K_{\eps\eta}}{G_u}c_{\eps\eta}\frac{ua_{jk}K_{\eps\eta}}{G_u}
-\int\! Bc a_{ki}c\frac{K_\eps^2}{K_{\eps\eta}^2}c
\frac{ua_{ij}K_{\eps\eta}}{G_u}c_{\eps\eta}\frac{ua_{jk}K_{\eps\eta}}{G_u}
\right.\nn\\
&\left.\quad
+\int\! Bc a_{ki}c_{\eps\eta}\frac{ua_{ij}K_{\eps\eta}}{G_u}c\frac{K_\eps^2}{K_{\eps\eta}^2}c
\frac{ua_{jk}K_{\eps\eta}}{G_u}
\r]
\nn\\
&\quad
-\sum_{i,j}\int_0^1\!\! du\l[
\int\! ca_{ij}c_{\eps\eta}K_{\eps\eta}c_{\eps\eta}\frac{ua_{ji}K_{\eps\eta}}{G_u}\l(1-\frac{1}{G_u}\r)
+\int\! Bca_{ij}c\frac{K_\eps^2}{K_{\eps\eta}^2}c \frac{ua_{ji}K_{\eps\eta}}{G_u}\l(1-\frac{1}{G_u}\r)cK
\r]
\nn\\
\calH_{\eps\eta}[G(K)]&=
\sum_{i,j,k}\int_0^1\!\! du\l[
\int\! Bc a_{ki}c_{\eps\eta}\frac{ua_{ij}K_{\eps\eta}}{G_u}c_{\eps\eta}
\frac{ua_{jk}K_{\eps\eta}}{G_u}cK_\eps^2
\right.
\nn\\
&\left.\quad
+\int\! Bca_{ki}cK_\eps^2\CR{c}{\frac{1}{K_{\eps\eta}^2}}
K_\eps^2 Bc
\frac{ua_{ij}K_{\eps\eta}}{G_u}c_{\eps\eta}\frac{ua_{jk}K_{\eps\eta}}{G_u}
\right.\nn\\
&\left.\quad
-\int\! Bca_{ki}c_{\eps\eta}\frac{ua_{ij}K_{\eps\eta}}{G_u}cK_\eps^2
\CR{c}{\frac{1}{K_{\eps\eta}^2}}
K_\eps^2Bc\frac{ua_{jk}K_{\eps\eta}}{G_u}
\r],
\nn\\
&\quad
-\sum_{i,j}\int_0^1\!\!du\l[
\int\! Bca_{ij}cK_\eps^2\CR{c}{\frac{1}{K_{\eps\eta}^2}}
K_\eps^2 Bc
 \frac{ua_{ji}K_{\eps\eta}}{G_u}\l(1-\frac{1}{G_u}\r)cK
 \right.\nn\\
 &\left.\quad
 -\int \!Bca_{ij} c_{\eps\eta}K_{\eps\eta}c_{\eps\eta}
 \frac{ua_{ji}K_{\eps\eta}}{G_u}\l(1-\frac{1}{G_u}\r)cK_\eps^2
 \r],
\end{align}
where all argument of $G,a_{ij}$ are $K_{\eps\eta}$.
We used the following relation to obtain $\calE_{\eps\eta},\calF_{\eps\eta}$.
\begin{align}
c_{\eps\eta}K_{\eps\eta}c_{\eps\eta}&=Qc_{\eps\eta}+\eta cK_\eps^2\CR{c}{\frac{1}{K_{\eps\eta}^2}}K_\eps^2 Bc
-\eps c\frac{K_\eps^2}{K_{\eps\eta}^2}c,
\nn\\
c_{\eps\eta}K_{\eps\eta}B_{\eps\eta}c_{\eps\eta}&=Q(Bc)+\eps c +\eta cK_\eps^2 Bc.
\end{align}
We define $\calE_{\eps}$ as $\calE_{\ee}$ with $\eta=0$;
\begin{align}
\calE_{\eps}[G(K),a_{ij}(K)]&=
\sum_{ijk}\int_0^1\!\! du\int \!ca_{ki}cK_\eps\CR{\frac{ua_{ij}}{G_u}}{c}\frac{ua_{jk}K_{\eps}}{G_u}.
\end{align}
Here arguments of $G,a_{ij}$ are $K_\eps$.
And we defined $\calH_{\eta}$ as $\calH_{\ee}$ with $\eps=0$, 
\begin{align}
\calH_{\eta}[G(K),a_{ij}(K)]&=
\int_0^1\!\! du\int Bc a_{ki}c_{\eta}\frac{ua_{ij}}{G_u}\frac{K}{1+\eta K}
c_{\eta}\frac{ua_{jk}}{G_u}
\frac{K}{1+\eta K}cK^2
\nn\\
&\quad
-\int_0^1\!\! du\int Bca_{ki}cK^2\CR{c}{\frac{(1+\eta K)^2}{K^2}}
K^2 Bc
\frac{ua_{ij}}{G_u}
\frac{K}{1+\eta K}
c_{\eta}\frac{ua_{jk}}{G_u}
\frac{K}{1+\eta K}
\nn\\
&\quad
+\int_0^1\!\! du\int Bca_{ki}c_{\eta}\frac{ua_{ij}}{G_u}
\frac{K}{1+\eta K}
cK^2
\CR{c}{\frac{(1+\eta K)^2}{K^2}}
K^2\frac{ua_{jk}}{G_u}
\frac{K}{1+\eta K}
\nn\\
&\quad
-\int_0^1\!\! du\int Bca_{ij}cK^2\CR{c}{\frac{(1+\eta K)^2}{K^2}}
K^2 Bc
 \frac{ua_{ji}}{G_u}\l(1-\frac{1}{G_u}\r)\frac{K}{1+\eta K}cK
 \nn\\
 &\quad
 -\int_0^1\!\! du\int Bca_{ij} c_{\eta}\frac{K}{1+\eta K}c_{\eta}
 \frac{ua_{ji}}{G_u}\frac{K}{1+\eta K}\l(1-\frac{1}{G_u}\r)cK^2,
\end{align}
where $c_\eta$ is defined as
\begin{align}
c_\eta=c(1+\eta K)^2 Bc.
\end{align}
Using the relations
\begin{alignat}{3}
&\CR{c}{\frac{(1+\eta K)^2}{K^2}}K^2Bc
&\quad
\overset{\text{inv}}{\to} 
&\quad
\CR{c}{(K+\eta)^2}Bc,
\nn\\
&c(1+\eta K)^2Bc 
&\quad
\overset{\text{inv}}{\to} 
&\quad c(K+\eta)^2Bc,
\nn\\
&c_{\eta}\frac{K}{1+\eta K} c_\eta &\quad\overset{\text{inv}}{\to }&\quad
c(K+\eta)\CR{c}{K}(K+\eta)Bc.
\end{alignat}
$\calH_{\eta}$ becomes
\begin{align}
\calH_{\eta}[G(K),a_{ij}(K)]&=
\int_0^1\!\! du\int c a_{ki}c(K+\eta)^2Bc\frac{ua_{ij}}{G_u(K+\eta)}c(K+\eta)^2Bc
\frac{ua_{jk}}{G_u(K+\eta)}
\nn\\
&\quad
+\int_0^1\!\! du\int Bca_{ki}cK^2Bc\frac{1}{K^2}\CR{c}{(K+\eta )^2} Bc
\frac{ua_{ij}}{G_u(K+\eta)}c\frac{ua_{jk}(K+\eta)}{G_u}
\nn\\
&\quad
-\int_0^1\!\! du\int Bca_{ki}c(K+\eta)^2Bc\frac{ua_{ij}}{G_u(K+\eta)}cK^2Bc \frac{1}{K^2}
\CR{c}{(K+\eta)^2}
\frac{ua_{jk}}{G_u(K+\eta)}
\nn\\
&\quad
-\int_0^1\!\! du\int Bca_{ij}cK^2Bc\frac{1}{K^2} \CR{c}{(K+\eta)^2}Bc\frac{ua_{ji}}{G_u^2(K+\eta)}(1-G_u)cK
\nn\\
&\quad
-\int_0^1\!\! du\int ca_{ij} c(K+\eta)\CR{c}{K}(K+\eta)Bc
\frac{ua_{ji}}{G_u(K+\eta)}\l(1-\frac{1}{G_u}\r)
\nn\\
&=\calE_\eta[G(1/K),a_{ij}(1/K)]
+\int Q\l(\int_0^1\!\! du Bca_{ij}c(K+\eta)^2 c\frac{ua_{ji}}{G_u^2(K+\eta)}(1-G_u)\r),
\end{align}
where $\calE_\eta$ is defined as replacing $\eps$ with $\eta$ in $\calE_{\eps}$.
The integration form of the regularized $Q$-exact quantity in the last line 
safely vanish.
Note that the arguments of $G$ and $a_{ij}$ 
changed from $ K/(1+ \eta K) $ to $1/(K+ \eta)$ by the inversion map Eq.\eqref{eq:inv_map}.

Concrete expression of $\calE'_{\ee}$ and $\calH'_{\ee}$ are given by
\begin{align}
\calE'_{\eps\eta}[G(K),a_{ij}(K)]&=\sum_{i,j,k,\ell}
\int Bca_{jk}(K_{\ee})c\frac{K_\eps^2}{K_{\eps\eta}}
\left(\delta_{k\ell}+\frac{a_{k\ell}}{G}(K_\ee)\right)
ca_{\ell i}(K_{\ee})c \left(\delta_{ij}+\frac{a_{ij}}{G}(K_{\ee})\right)
\frac{K_\eps^2}{K_{\eps\eta}}
\nn\\
\calH'_{\eps\eta}[G(K),a_{ij}(K)]&=
\sum_{i,j,k,\ell}\int Bc a_{jk}(K_{\ee})cK_\eps^2
\CR{\frac{1}{K_{\ee}}\left(\delta_{k\ell}+\frac{a_{k\ell}}{G}(K_{\ee})\right)}{c}K_\eps^2
\nn\\
&\quad\times 
\CR{a_{\ell i}(K_{\ee})}{c}\left(\delta_{ij}+\frac{a_{ij}}{G}(K_{\ee})\right)
\frac{K_\eps^2}{K_{\eps\eta}}
\end{align}
where all arguments of $G$ and $a_{ij}$ are $K_{\eps\eta}$.
And we define $\calE'_{\eps}$($\calH'_\eta$) as $\calE'_{\ee}$($\calH'_{\ee}$) with $\eta=0$($\eps=0$).
\begin{align}
\calE'_{\eps}[G(K),a_{ij}(K)]&=
\sum_{i,j,k,\ell}
\int Bca_{jk}(K_{\eps})cK_\eps
\left(\delta_{k\ell}+\frac{a_{k\ell}}{G}(K_\eps)\right)
ca_{\ell i}(K_{\eps})c \left(\delta_{ij}+\frac{a_{ij}}{G}(K_{\eps})\right)K_\eps,
\nn\\
\calH'_{\eta}[G(K),a_{ij}(K)]&=
\sum_{i,j,k,\ell}\int Bc a_{jk}(K_{\eta})cK^2Bc
\frac{1+\eta K}{K}\left(\delta_{k\ell}+\frac{a_{k\ell}}{G}(K_{\eta})\right)cK^2Bc
\nn\\
&\quad\times 
a_{\ell i}(K_{\eta})cK^2Bc\left(\delta_{ij}+\frac{a_{ij}}{G}(K_{\eta})\right)
\frac{1+\eta K}{K},
\end{align}
where all argument of $G$ and $a_{ij}$ are $K_\eps$($K/(1+\eta K)$) in 
$\calE'_\eps$ ($\calH'_\eta$).
Just as in the above discussion, we can prove that the following relation exists between $\calE'$ and $\calH'$.
\begin{align}
\calH'_\eta[G(K),a_{ij}(K)]=\calE'_\eps[G(1/K),a_{ij}(1/K)],
\end{align}
where $\calE'_\eta$ is defined as replacing $\eps$ with $\eta$ in $\calE'_{\eps}$.

\providecommand{\href}[2]{#2}\begingroup\raggedright\endgroup

\end{document}